\documentclass[aps,twocolumn,showpacs,amsmath,amssymb,prl,superscriptaddress]{revtex4}

\usepackage{graphicx}
\usepackage{dcolumn}
\usepackage{bm}
\usepackage{epsf}
\usepackage{amsmath}
\usepackage{amssymb}
\usepackage{color}

\newcommand{\tr}[1]{\mathrm{tr}\left\{#1\right\}}

\newcommand{\la}{\left\langle}
\newcommand{\ra}{\right\rangle}

\newcommand{\bla}{bla\\bla\\bla\\bla\\bla}

\newcommand{\PRE}{Phys. Rev. E }
\newcommand{\PRL}{Phys. Rev. Lett. }
\newcommand{\EPL}{Europhys. Lett. }

\begin{document}

\title{
Efficiency of heat engines coupled to nonequilibrium reservoirs}

\author{Obinna Abah}

\affiliation{Department of Physics, University of Augsburg, D-86135 Augsburg, Germany}
\affiliation{Dahlem Center for Complex Quantum Systems, FU Berlin, D-14195 Berlin, Germany}
\author{Eric Lutz}
\affiliation{Dahlem Center for Complex Quantum Systems, FU  Berlin, D-14195 Berlin, Germany}
\affiliation{Institute for Theoretical Physics, University of Erlangen-N\"urnberg, D-91058 Erlangen, Germany}

\pacs{05.30.-d, 03.65.-w}

\begin{abstract} 
We consider quantum heat engines that operate between  nonequilibrium stationary reservoirs.
We evaluate their maximum efficiency from the positivity of the entropy production and show that it can be expressed in terms of an effective temperature that depends on the nature of the reservoirs. We further compute the efficiency at maximum power for different kinds of engineered reservoirs and derive a nonequilibrium generalization of the Clausius statement of the second law.
   
\end{abstract}

\maketitle

Engines are devices that convert various forms of energy into useful mechanical work and motion. In thermodynamics, two different kinds of machines can be distinguished. On the one hand, there are heat engines that operate between two reservoirs at different temperatures, such as internal combustion engines \cite{cal85,cen01}. On the other hand, there are molecular motors that are driven  from equilibrium by varying external parameters,  while  in contact with a single isothermal reservoir \cite{jul97,rei02}. The latter describe biological motor proteins as well as artificial nanomachines \cite{kay07,han09}.  An essential characteristic of any machine is its efficiency defined as the ratio of  work output to energy input. Whereas for heat engines the efficiency is limited by the  Carnot formula, $\eta_c= 1-T_1/T_2$, where $T_1$ and $T_2$ are the temperatures of the two thermal reservoirs ($T_1<T_2$), it can reach unity for molecular motors \cite{par99,ast07,toy10,toy11}. Maximum efficiency usually corresponds to quasistatic conditions, and therefore to zero power. A practically more relevant quantity is thus the efficiency at maximum power which for heat engines is given by $\eta_c/2$ for small temperature differences \cite{sch08,esp09,esp10}. For molecular motors, the efficiency at maximum power  can  reach the thermodynamic limit 1 for strong driving \cite{sei11,bro12}.

Heat engines are usually assumed to be in contact  with two equilibrium reservoirs. In this paper, we investigate the more general case where the engine runs between  stationary nonequilibrium reservoirs.   
In a sense, this situation interpolates between  traditional heat engines and  molecular motors. Indeed, the efficiency of these heat engines may be larger than the Carnot efficiency and  they may operate  isothermally. Our study is motivated by the recent advent of reservoir engineering techniques in quantum optical systems, such as ion traps \cite{poy96,mya00}, microwave cavities \cite{pie07,sar11}, optical lattices \cite{die08,ver09} and optomechanical systems \cite{wan13}, that enable the preparation of  nonthermal environments.  In addition, theoretical studies have shown in individual cases that the efficiency of heat engines coupled to nonthermal quantum coherent \cite{scu03}  or quantum correlated \cite{dil09} reservoirs may sometimes exceed the Carnot value. The two fundamental questions that we here  address are therefore: i) what is the maximum (universal) efficiency that may be reached, and ii) under what conditions is this  efficiency larger  than the standard Carnot limit? In the following, we consider a quantum heat engine coupled to   general stationary nonthermal  reservoirs. It will be convenient to regard these reservoirs as perturbed thermal reservoirs. They will then be characterized by a temperature and a second  parameter (or more) that quantifies the deviation from equilibrium, such as the degree of quantum  coherence \cite{scu03} or the amount of  quantum correlations \cite{dil09}.  We begin by performing a detailed analysis of the quantum Otto cycle for a time-dependent harmonic oscillator,  a paradigm of quantum heat engines and a generalization of the common four-stroke car engine \cite{scu02,lin03,rez06,qua07,aba12}. We evaluate its efficiency which we express in terms of the Hamiltonian of mean force \cite{cam09,hil11}, a quantum extension of the potential of mean force known in the statistical theory of fluids \cite{hil56}. We  derive  an explicit expression for the maximum efficiency of a heat engine from the condition of positive entropy production  of the second law of thermodynamics \cite{jar11}. This efficiency may be larger or smaller than the Carnot efficiency depending on the properties of the reservoirs, which we quantify with an effective temperature.  We further obtain a generalization of the Clausius statement of the second law on the direction of heat flow between nonequilibrium systems characterized by their effective temperatures. Finally, we compute  the efficiency at maximum power of the Otto engine for the concrete examples  of  correlated    and coherent quantum reservoirs. 

\textit{Quantum Otto engine.} We consider a quantum Otto engine whose working medium is a harmonic oscillator with time-dependent frequency $\omega_t$ \cite{scu02,lin03,rez06,qua07,aba12}. The Otto cycle consists of two isentropic processes during which the frequency  is unitarily varied between $\omega_1$ and $\omega_2$, and of two isochoric (constant frequency) processes during which the oscillator is connected to two different reservoirs (see Fig.~1). A concrete scheme to experimentally realize such an engine using a single ion in a linear Paul trap has been proposed in Ref.~\cite{aba12}. In the usual Otto cycle, the two reservoirs are assumed to be thermal and characterized by the  inverse temperatures $\beta_i = 1/(k T_i)$, ($i=1,2$), where $k$ is the Boltzmann constant. Here, we examine the situation where the engine is alternatingly coupled to  engineered nonthermal reservoirs. For simplicity, we will first focus on the case where only the hot reservoir is nonthermal. When connected to this  reservoir, the oscillator relaxes to a nonequilibrium state that we write in the general form, $\rho_2= \exp[-\beta_2 (H_2 + \Delta H_2)]/Z_2^*$, where $H_2 = p^2/(2m) + m\omega_2^2 x^2/2$ is  the  Hamiltonian of the oscillator at frequency $\omega_2$ ($m$ denotes the mass) and  $Z_2^*$ the normalization constant. The operator $\Delta H_2$ quantifies the departure from equilibrium and may be arbitrary \cite{rem0,rem3}.  We note that   the operator $H_2^*= H_2 + \Delta H_2$ can be seen as a Hamiltonian of mean force \cite{cam09,hil11}.  It is important to stress that the  state $\rho_2$ should be centered, $\la x\ra = \la p\ra =0$, so  that the nonequilibrium reservoir is a proper heat source that can only exchange heat with the engine, but no work \cite{fer56}.  The mean energy of the harmonic oscillator in the nonequilibrium state is then,
 \begin{equation}
\label{1}
\langle H_2 \rangle =  \hbar \omega_2 (\bar n_2+ \Delta \bar n + 1/2),
\end{equation}
where $\bar n_i = [\exp(\hbar \beta_i\omega_i)-1]^{-1}$ is the mean  occupation number of a thermal quantum oscillator and $ \Delta \bar n=\Delta \bar n(\omega_2)$ that associated with the deviation from the thermal state. The density operator of the harmonic oscillator in contact with the cold thermal reservoir is $\rho_1= \exp(-\beta_1 H_1)/Z_1$ with mean energy $ \la H_1\ra= \hbar \omega_1 (\bar n_1+1/2)$. 

\begin{figure}[t]
\includegraphics[width=\columnwidth]{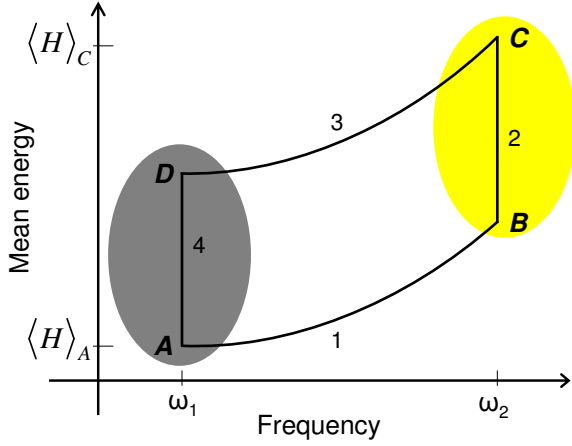}
\caption{(Color online) Energy-frequency diagram of the quantum Otto engine. The thermodynamic cycle consists of two isentropic (stroke 1 and 3) and  two isochoric (stroke 2 and 3) processes. In the latter, the heat engine is coupled to engineered nonthermal stationary reservoirs that are described by a temperature and additional parameters that characterize the deviation from thermal equilibrium.\label{cycle}}
\end{figure}

During the two isentropic parts of the thermodynamic cycle (stroke 1 and 3), the time-dependent oscillator is isolated and its dynamics is hence unitary. As a result, its Schr\"odinger equation can be solved exactly using a Gaussian wave function ansatz and the mean energy can be evaluated analytically \cite{hus53,def08,def10}. During the two isochoric branches (stroke 2 and 4), the oscillator relaxes respectively to a nonequilibrium and an equilibrium state. The corresponding average energies of the harmonic oscillator at the four corners of the quantum Otto cycle are,
\begin{subequations}
\begin{eqnarray}
\la H\ra_{A}&=&\hbar\omega_1 (\bar n_1 + 1/2),\\
\la H\ra_{B}&=&\hbar\omega_2\,Q^\ast_1(\bar n_1 + 1/2),\\
\la H\ra_{C}&=&\hbar\omega_2\,(\bar n_2+  \Delta \bar n + 1/2),\\
\la H\ra_{D}&=&\hbar\omega_1\,Q^\ast_2 (\bar n_2 +   \Delta \bar n + 1/2),
\end{eqnarray}
\end{subequations}
where we have used Eq.~\eqref{1}  and Eq.~(5.12) of Ref.~\cite{hus53}. We note that the energies at point $C$ and $D$ are modified by the presence of the nonthermal reservoir. The two parameters $Q^\ast_1$ and $Q^\ast_2$ characterize the degree of adiabaticity of the compression and expansion phases 1 and 3 \cite{hus53}.  Their explicit expressions  for any given frequency modulation $\omega_t$, can be found in Refs. \cite{def08,def10}. They are, for example,  equal to one for adiabatic processes and to $(\omega_1^2+\omega_2^2)/(2\omega_1\omega_2)$ for a sudden frequency change. 

To evaluate the efficiency of the Otto engine, we need to compute work and heat  along  the four branches of the cycle (see Fig.~1).
The mean works, denoted by $\la W_1\ra$ and  $\la W_3\ra$, done during  stroke 1 and 3 are given by,
\begin{eqnarray}
\label{3}
\la W_1\ra&=&\la H\ra_{B}-\la H\ra_{A}
=(\hbar\, \omega_2\,Q^\ast_1 - \hbar\, \omega_1)(\bar n_1 + \frac{1}{2}),\\
\la W_3\ra &=& \la H\ra_D-\la H\ra_C 
= (\hbar\omega_1\,Q^\ast_2 - \hbar\omega_2)(\bar n_2 +   \Delta \bar n + \frac{1}{2}). \nonumber
\end{eqnarray}
At the same time, the mean heats, $\la Q_2 \ra$ and $\la Q_4 \ra$, exchanged with the reservoirs during stroke 2 and 4 read,
\begin{eqnarray}
\label{4}
\la Q_2\ra&=&\la H\ra_C-\la H\ra_B \nonumber\\
&=&\hbar\omega_2 (\bar n_2+ \Delta \bar n + \frac{1}{2}) -\hbar\omega_2 Q^\ast_1(\bar n_1 + \frac{1}{2}),\\
\la Q_4\ra&=&\la H\ra_A-\la H\ra_D \nonumber \\
&=&\hbar\omega_1 (\bar n_1+ \frac{1}{2})  - \hbar\omega_1 Q^\ast_2(\bar n_2 +  \Delta \bar n + \frac{1}{2}).\nonumber
\end{eqnarray}
 The efficiency, defined as the ratio of the total work per cycle, $\la W\ra = - (\la W_1\ra + \la W_3\ra)$, to the heat received from the hot reservoir, $\la Q_2\ra$, can then be written  as,
\begin{eqnarray}
\label{5}
 \eta =1-\frac{\omega_1}{\omega_2}\,\frac{(\bar n_1+ 1/2) - Q^\ast_2\,(\bar n_2+  \Delta \bar n + 1/2)}{Q^\ast_1\,(\bar n_1 + 1/2)-(\bar n_2 +  \Delta \bar n + 1/2)}.
\end{eqnarray}
The above quantum expression is exact. It gives the finite time efficiency of the quantum Otto engine for any frequency modulation $\omega_t$, any inverse temperature $\beta_i$, and any nonequilibrium stationary reservoir. 

\textit{Maximum efficiency.} 
The maximum efficiency  can be evaluated from the
 positivity of the entropy production   \cite{cal85,ali79}. Applying the Klein inequality, $S(\rho_r||\rho_s)\geq0 $ \cite{nie00}, to the isochoric processes $BC$ and $DA$, we obtain,
\begin{eqnarray}
\label{60}
 &&\hspace{-.5cm}S(\rho_B||\rho_C) + S(\rho_D||\rho_A)\nonumber \\
 &&\hspace{-.5cm}= -\beta_2\la Q_2\ra -\beta_1 \la Q_4\ra +\beta_2 \tr{(\rho_B-\rho_C) \Delta H_2} \geq 0,
\end{eqnarray}
where $\rho_r$ is the density operator of  state $r$ and the quantum relative entropy, $S(\rho_r||\rho_s) = \tr{\rho_r(\ln \rho_r-\ln \rho_s)}$,  the entropy production associated with the thermalization step $r\rightarrow s$   \cite{def11}. We note that the von Neumann entropy remains constant during the  isentropic  processes $AB$ and $CD$. To evaluate the last term in Eq.~\eqref{60}, we use  that $\la \Delta H\ra_r=\tr{\rho_r \Delta H_2}= (a/2) \la p^2\ra_r + (b/2) \la x^2\ra_r$, where the two coefficients $a$ and $b$ are given by \cite{hil11},
\begin{equation}
\label{61}
a= \frac{1}{\beta_2\la p^2\ra_C} - \frac{1}{m}, \qquad b=\frac{1}{\beta_2\la x^2\ra_C} - m\omega_2^2,
\end{equation}
in the high-temperature limit, $\bar n_i +1/2\simeq 1/(\hbar \beta_i\omega_i)$. For small deviations from equilibrium, momentum and position quadratures at point $C$ can be written as 
$\la p^2\ra_C \simeq m\hbar \omega_2 (\bar n_2 + \Delta \bar n)$ and $\la x^2\ra_C\simeq (\hbar/m\omega_2) (\bar n_2 + \Delta \bar n)$. As a result,  for adiabatic frequency modulation, $Q^*_{1,2} =1$, which corresponds to maximum efficiency,
\begin{eqnarray}
\label{62}
\la \Delta H\ra_B&=&kT_1 \frac{\omega_2}{\omega_1}\left(\frac{T_2}{T_2^\text{eff}}-1\right),\nonumber \\ \la \Delta H\ra_C&=& k(T_2-T_2^\text{eff}).
\end{eqnarray}
In the above equations, we have introduced  the effective temperature, $T_2^\text{eff}= T_2+ \hbar \omega_2 \Delta \bar n/k= T_2 +\la \Delta H\ra_C/k$, that quantifies the departure from equilibrium of the hot nonthermal reservoir. For adiabatic frequency modulation and high temperature,  Eq.~\eqref{4} can be expressed in the form, $\la Q_2\ra= k T_2^\text{eff} -kT_1 \omega_2/\omega_1$. Hence $-\beta_2 \la Q_2\ra +\beta_2 (\la \Delta H\ra_B-\la \Delta H\ra_C) = -\beta_2^\text{eff} \la Q_2\ra$, and the second law of thermodynamics yields the inequality,
\begin{equation}
\label{63}
-\beta_2^\text{eff} \la Q_2\ra -\beta_1 \la Q_4\ra \geq 0.
 \end{equation}
 On the other hand, according to the first law,
 \begin{equation}
 \label{64}
\la W\ra =  \la Q_2\ra + \la Q_4\ra .
 \end{equation}
 Combining Eqs.~\eqref{63} and \eqref{64}, we eventually find $\la Q_2\ra (\beta_1-\beta_2^\text{eff}) \geq \beta_1 \la W\ra$. We can therefore  conclude that the maximum efficiency of the engine is,
  \begin{equation}
  \label{65}
\eta_\text{max}=1-\frac{\beta_2^\text{eff}}{\beta_1}= 1-\frac{\beta_2}{\beta_1(1+\beta_2 \la \Delta H\ra_C)}.
 \end{equation}
 We note  that since the work produced by the engine is positive,  the heat $\la Q_2\ra \geq 0$  only if $T_2^\text{eff}\geq T_1$, that is, heat flows from the high (effective) temperature to the low temperature reservoir. In particular, in the absence of the heat engine, $\la W \ra =0$, heat will flow from the system at high effective temperature to the system at low effective temperature \cite{rem2}. 
We have thus generalized  the  Clausius statement of the second law \cite{cla79}  to stationary systems that  are not in thermal equilibrium, extending analogous equilibrium derivations found e.g. in Refs.~\cite{tol38,par89}. 

The  Clausius principle provides alternative means to derive the maximum efficiency \cite{fer56}. Assuming that  
$T_2^\text{eff}\geq T_1$, we observe that heat is  absorbed from the hot reservoir, $\la Q_2\ra \geq 0$, and flows into the cold reservoir, $\la Q_4\ra \leq 0$. Equation \eqref{4} then leads to,
\begin{equation}
\label{6}
\frac{\bar n_2+ \Delta \bar n +1/2}{\bar n_1 + 1/2}\geq Q_1^*, \quad \frac{\bar n_1 +1/2}{\bar n_2 + \Delta \bar n + 1/2}\leq Q_2^*.
\end{equation}
 Combining Eqs.~\eqref{5} and \eqref{6} for adiabatic frequency modulation, $Q^*_{1,2}=1$, and high temperatures, we recover the maximum efficiency \eqref{65}.
 
 Equation \eqref{65} generalizes the standard Carnot formula to heat engines that operate between a thermal and a nonthermal stationary reservoir. It only depends on  the two temperatures of the reservoirs and on the energy deviation from equilibrium at point $C$, $\la\Delta H\ra_C=\la H\ra_C - \la H\ra^\text{eq}_C=\hbar\omega_2  \Delta \bar n$.   Expression \eqref{65} exceeds the Carnot efficiency $\eta_c$ when  $\la \Delta H\ra_C>0$.  This   condition corresponds to a larger area of the thermodynamic cycle  in Fig.~1, and therefore to a larger work output. Equation \eqref{65} further indicates that the nonthermal reservoir can be described by an effective temperature, $T_2^\text{eff}$,  that may be larger or smaller than that of the unperturbed  thermal reservoir (see below). This situation is reminiscent of that of molecular motors where the external  driving is sometimes regarded as a nonequilibrium reservoir with an effective temperature (see e.g. Ref.~\cite{rei02}, Sect.~3.4.2).  
 
 Expression \eqref{65} agrees with the maximum efficiency obtained for a  quantum Carnot engine in contact with either a quantum coherent \cite{scu03} or a  quantum correlated \cite{dil09}  reservoir  (see below). We have here generalized these results to quantum nonthermal reservoirs of arbitrary nature. We note, moreover,  that the fact that Eq.~\eqref{65} appears as the maximum efficiency of different kinds of quantum heat engines (Otto and particularly Carnot) strongly hints  at its universal validity for all heat engines.
 
An important observation is that in thermodynamics the two thermal reservoirs are supposed to be given  \cite{cal85,cen01}. In particular, the energetic cost of preparing, say, a high temperature reservoir in addition to an ambient low temperature reservoir is not taken into account in the calculation of the efficiency of a heat engine  \cite{rem1}. Such an inclusion would indeed lead to vanishing efficiencies due to the large (strictly speaking infinite) energy content of a proper heat reservoir.  We  here follow the same approach and consider the nonthermal reservoir as given. In this framework, the Carnot formula appears as an expression of the second law of thermodynamics for a specific form of nonequilibrium (two thermal reservoirs at different temperatures), whereas the efficiency \eqref{65} applies to a  more general form of nonequilibrium (one thermal and one nonthermal reservoir). In a similar manner, the maximum  efficiency  of 1 of  molecular motors is a consequence of the second law for this yet different type of nonequilibrium \cite{par02}. It is worth noticing that Eq.~\eqref{65} yields a non-zero result, $\eta_\text{max}= \beta_2 \la\Delta H\ra_C/(1+\beta_2 \la\Delta H\ra_C)$, for isothermal reservoirs, $\beta_1=\beta_2$. In this situation, which is akin to that of molecular motors, the nonthermal reservoir can be seen as an external nonequilibrium driving.

\textit{Efficiency at maximum power.} The efficiency at maximum power is often a more relevant quantity than the maximum efficiency which  corresponds to zero power \cite{cur75}. In contrast to the latter, however, there does not seem to be a universal expression for the efficiency at maximum power; it not only usually depends on the optimization procedure, but also, as we will show, on the details of the nonthermal reservoir. The power output of a heat engine is defined as $P= \la W\ra /\tau$, where $\tau$ is the duration of the cycle. In the following, we evaluate the maximum efficiency and the efficiency at maximum power for two different examples. For simplicity, we focus on adiabatic compression and expansion,  $Q^\ast_{1,2} = 1$, since nonadiabatic processes lead to smaller efficiencies \cite{rez06,aba12}, and to the high-temperature regime $\beta_i\hbar\omega_i \ll 1$.

 Let us consider a quantum photo-engine made of a single mode (the harmonic oscillator)  in a resonant cavity with moving mirror, and coupled to a beam of thermal two-level atoms that pass through the cavity \cite{scu03,dil09}. When the atoms are uncorrelated, the beam plays the role of a thermal reservoir. By contrast, for correlated atoms the engineered reservoir is nonthermal. For  pairwise thermally entangled atoms, the deviation of the mean occupation number from equilibrium is  \cite{dil09},
\begin{equation}
\Delta \bar n_{\lambda:1} = \frac{\beta_2\hbar\lambda^2}{4\omega_2}, \qquad
\Delta \bar n_{\lambda:2} = -\frac{\lambda}{2\omega_2},
\end{equation}
when respectively one or the two atoms of a correlated pair fly through the cavity, in the limit of high temperature, $\beta_i\hbar\omega_i \ll 1$, and weak correlation, $\beta_i\hbar\lambda \ll 1$. Here $\lambda$ is the strength of the interaction that created the thermal entangled pair. The maximum efficiency \eqref{65} is larger than the Carnot expression when one atom o a pair flies through the cavity $(\Delta \bar n_{\lambda:1}>0)$ and smaller when the two atoms of a pair  pass through it $(\Delta \bar n_{\lambda:2}<0)$. In both cases, the deviation of the mean occupation number is inversely proportional to the frequency $\omega_2$. The total work produced by the engine  during one cycle is,
\begin{equation}
-\la W\ra = \frac{1}{\beta_1}\left(\frac{\omega_2}{\omega_1} - 1\right) + \left(\frac{1}{\beta_2} + \hbar \omega_2 \Delta \bar n_{\lambda} \right)\left(\frac{\omega_1}{\omega_2} - 1\right).
\end{equation}
 Assuming that the initial frequency of the oscillator $\omega_1$ (as well as $\lambda,\,\beta_1\, ,\beta_2$ and the cycle time) are fixed and by optimizing with respect to the second frequency $\omega_2$, we find that the power is maximum  when 
$\omega_1/\omega_2 = \sqrt{\beta_2/(\beta_1 [1+\hbar \beta_2 \omega_2 \Delta \bar n_{\lambda}])}$.
As a result, the efficiency at maximum power is given by,
\begin{equation}
\label{10}
\eta_\gamma = 1 - \sqrt{\frac{\beta_2}{\beta_1 \left(1 + \beta_2 \la \Delta H\ra_C\right)}} = 1 -\sqrt{\frac{\beta_2^\text{eff}}{\beta_1}}.
\end{equation} 
Equation \eqref{10} reduces to the Curzon-Ahlborn expression \cite{cur75} for vanishing correlation and generally exceeds it when  one  atom flies through the cavity, $\la \Delta H\ra_C=\hbar \omega_2 \Delta \bar n_{\lambda}>0$. Remarkably, Eq.~\eqref{10} can be expressed in terms of the same effective temperature as Eq.~\eqref{65}.  This remains  true for all reservoirs with $\Delta \bar n(\omega_2) \sim 1/\omega_2$. A discussion of the efficiency at maximum power for a quantum coherent reservoir \cite{scu03}, for which $\Delta \bar n(\omega_2) \sim 1/\omega_2^2$, is presented in the Appendix. In the limit of small temperature differences and small $\la \Delta H\ra_C$, we  have $\eta_\gamma \simeq \eta_c/2 + \beta_2^2 \la \Delta H\ra_C/(2\beta_1) \geq \eta_c/2$, when $\la \Delta H\ra_C>0$; this   result therefore  lies beyond the range of the usual linear regime \cite{sch08,esp09,esp10}.

\textit{Generalization.} The above results can be  extended to situations where the two reservoirs are nonthermal. For instance, the efficiency of the  quantum Otto engine is,
\begin{equation}
\label{16}
\eta = 1 - \frac{\omega_1}{\omega_2}\frac{(\bar n_1 + \Delta \bar n_1 + 1/2) -Q^{\ast}_2(\bar n_2 + \Delta \bar n_2 + 1/2)}{(\bar n_1 + \Delta \bar n_1 + 1/2)Q^{\ast}_1 - (\bar n_2 + \Delta \bar n_2 + 1/2)},
\end{equation}
where $\Delta \bar n_i$  is the deviation of the  mean occupation number for the nonthermal reservoir $i$ $(i=1,2)$. The positivity of the entropy production for this nonequilibrium configuration leads to the following  expression of the high-temperature maximum efficiency,
\begin{equation}
\label{17}
\eta_\text{max}=  1- \frac{\beta_2^\text{eff}}{\beta_1^\text{eff}}
=\eta_c + \beta_2 \left(\frac{\omega_1}{\omega_2}\la\Delta H\ra_C -\la\Delta H\ra_A\right),
\end{equation}
with the effective temperatures $T_2^\text{eff}= T_2+ \hbar \omega_2 \Delta \bar n_2/k= T_2 +\la \Delta H\ra_C/k$, as before, and $T_1^\text{eff}= T_1+ \hbar \omega_1 \Delta \bar n_1/k= T_2 +\la \Delta H\ra_A/k$. Here we have defined  the deviation,  $\la\Delta H\ra_A = \la H\ra_A - \la H\ra^\text{eq}_A$, of the energy of the oscillator from its equilibrium value at point $A$. The efficiency \eqref{17} is larger than the Carnot efficiency when $\omega_1\la\Delta H\ra_C - \omega_2\la\Delta H\ra_A >0$. It is interesting to note that  Eq.~\eqref{17}   only depends  on the average deviations at point $A$ and $C$ and not at points $B$ or $D$.

\textit{Conclusions.} We have used the  positivity of the entropy production to compute the maximum efficiency of a quantum heat engine operating between nonthermal stationary reservoirs. We have shown that the latter can be expressed in terms of an effective temperature which characterizes the deviation from equilibrium.   We have  obtained explicit conditions under which   the maximum efficiency exceeds  the standard Carnot bound in the presence of either one or two nonthermal reservoirs.  We have further derived a  generalization of the Clausius statement of the second law on  the direction of heat flow between two nonequilibrium systems. Additionally, we have  evaluated the efficiency at maximum power of the Otto engine for a quantum correlated reservoir  and obtained nonlinear extensions of the Curzon-Ahlborn formula. The efficiency of molecular motors has been experimentally shown to approach unity in some cases \cite{toy10,toy11}, far surpassing the efficiency of heat engines and highlighting the advantage of operating away from equilibrium. Our results provide a theoretical framework for a new class of engineered heat engines that interpolate between standard heat engines and molecular motors.

This work was supported by the DFG (contract No LU1382/4-1) and the Alexander von Humboldt Foundation.

\section{Appendix}
 A quantum coherent nonthermal reservoir can be created by sending through the optical cavity a beam of thermal three-level atoms whose degenerate ground states are prepared in a coherent superposition with relative phase $\phi$. In the high-temperature limit, the deviation of the mean occupation number is \cite{scu03},
\begin{equation}
\Delta \bar n_\phi = -\frac{1}{(\hbar\beta_2\omega_2)^2}\, \varepsilon\cos\phi,
\end{equation}
where $\varepsilon$ is proportional to the amplitude of the atomic coherence. Here $\Delta \bar n$ is inversely proportional to the square of the frequency $\omega_2$. The total work produced by the heat engine can be readily  written as,
\begin{equation}
-\la W\ra  = \frac{1}{\beta_1}\left(\frac{\omega_2}{\omega_1}-1\right)+ \frac{1}{\beta_2}\left(\frac{\omega_1}{\omega_2}-1\right)\left(1-\frac{\varepsilon\cos\phi}{\hbar\beta_2\omega_2}\right)
\end{equation}
By maximizing the power  with respect to $\omega_2$, keeping all other parameters constant as done in the main text, we find that the  power  is maximum when the following condition is satisfied, 
\begin{equation}
\frac{\omega_2}{\omega_1} = \sqrt{\frac{\beta_2}{\beta_1}} \left[1 - \frac{\varepsilon\cos\phi}{2\hbar\beta_2\omega_1}\left( 1- 2\, \sqrt{\frac{\beta_1}{\beta_2}}\right) \right],
\end{equation}
in a perturbation expansion for small values of $\varepsilon$ \cite{ben78}. The resulting efficiency at maximum power is then 
\begin{equation}
\label{15}
\eta_{\phi} = 1 - \sqrt{\frac{\beta_2}{\beta_1}}\, \left[1 - \frac{\varepsilon \cos\phi}{2 \hbar\beta_2\omega_1}\left( 1- 2\, \sqrt{\frac{\beta_1}{\beta_2}}\right) \right].
\end{equation}
The maximum efficiency, $\eta_\text{max}=1-\beta_2^\text{eff}/\beta_1$, and the efficiency at maximum power \eqref{15} exceed their thermal counterparts when the condition $\cos\phi<0$ is satisfied.


\begin{thebibliography}{99}
\bibitem{rem0} For any nonnormalized nonequilibrium operator $\rho_2'$, we define the normalized operator $\rho_2= \rho_2'/\tr{\rho_2'}$. We then have $\Delta H_2 = -[\ln \rho_2 + \ln \tr{\rho_2'}]/\beta_2 - H_2$.
\bibitem{rem3} A nonzero $\Delta H_2$ may lead to nonthermally distributed population and/or  nondiagonal energy matrix elements of $\rho_2$, as in the example of the coherent reservoir \cite{scu03}.
\bibitem{rem2} The derivation can be easily extended to two nonequilibrium systems: the two temperatures will be replaced by the corresponding effective temperatures.
\bibitem{rem1} In a similar way, the evaluation of the efficiency of a car engine does not include the energetic cost of extracting, refining and transporting  oil to the gas station.
\end{thebibliography}
\end{document}